\documentclass{tlp}

\renewcommand{\tt}{\usefont{OT1}{cmtt}{m}{n}\selectfont}
\usepackage{latexsym}
\usepackage{xspace}
\usepackage{graphicx}
\usepackage{url}

\makeatletter
\newenvironment{prog}{\ifhmode\par\vspace{1.2ex}\fi
\setlength{\parindent}{3ex}
\setlength{\parskip}{0.0ex}
\obeylines\@vobeyspaces\tt}{\vspace{1.2ex}\noindent
}
\makeatother
\newcommand{\startprog}{\begin{prog}}
\newcommand{\stopprog}{\end{prog}\noindent}
\newcommand{\code}[1]{\mbox{\tt #1}}   
\newcommand{\ccode}[1]{``\mbox{\tt #1}''}  
\newcommand{\bs}{\char92} 
\newcommand{\us}{\char95} 

\submitted{16 June 2010}
\revised{18 October 2010}
\accepted{31 July 2012}

\begin{document}
\sloppy

\title[An ER-based Framework for Declarative Web Programming]
 {An ER-based Framework for\\ Declarative Web Programming}

\author[M. Hanus and S. Koschnicke]
       {MICHAEL HANUS and SVEN KOSCHNICKE%
  \thanks{This work was partially supported by the
   German Research Council (DFG) under grant Ha 2457/5-2.}\\
        Institut f\"ur Informatik, CAU Kiel, D-24098 Kiel, Germany \\
        \email{mh@informatik.uni-kiel.de~~sven@koschnicke.de}}

\maketitle

\begin{abstract}
We describe a framework to support the implementation
of web-based systems intended to manipulate data stored in relational
databases. Since the conceptual model of a relational database
is often specified as an entity-relationship (ER) model,
we propose to use the ER model to generate
a complete implementation in the declarative programming language
Curry. This implementation contains operations to create and
manipulate entities of the data model, supports authentication,
authorization, session handling, and the composition of
individual operations to user processes.
Furthermore, the implementation ensures
the consistency of the database w.r.t.\ the data dependencies
specified in the ER model, i.e., updates initiated by the user
cannot lead to an inconsistent state of the database.
In order to generate a high-level declarative implementation
that can be easily adapted to individual customer requirements,
the framework exploits previous works on declarative
database programming and web user interface construction in Curry.
\end{abstract}


\section{Introduction}

Many web applications are in essence interfaces
on top of standard web browsers to work with data
stored in databases.
Typically, clients can access or modify existing data
as well as insert new data. The use of standard web browsers
demands for access control,
e.g., users must be authenticated, the authentication must be stored in
a session across various web pages, the access to various parts of the data
must be authorized, etc.
These requirements make the implementation of such applications
a non-trivial and often error-prone task \cite{Huseby03}.
In order to support the programmer in the design and implementation
of such web-based applications, various \emph{web frameworks}
had been developed for different implementation languages.
For instance, the popular Ruby on Rails
framework\footnote{\tt\url{http://www.rubyonrails.org/}}
supports the implementation of web applications in the
object-oriented language Ruby.
An interesting idea of this framework to enable the quick construction
of an initial system, which can be stepwise modified or extended,
is \emph{scaffolding}, i.e., the code of an initial implementation
is generated from the data model.
This initial code gives the programmer a good idea how to
structure and organize the code of the system under development.

This paper is based on a similar idea but exploits declarative
programming to obtain a compact implementation that can be easily
adapted and provides reliability in various aspects
(type safety, database consistency, etc).
For this purpose, we use the declarative multi-paradigm language
Curry \cite{Hanus97POPL,Hanus12Curry}
as an implementation language and exploit previous works
on declarative database programming \cite{BrasselHanusMueller08PADL}
and declarative construction of web user interfaces
\cite{Hanus06PPDP,Hanus07PPDP}.
Although some features of Curry, such as logic variables or narrowing,
are not directly used here, we remark that these features
are essential in the previous works to enable high-level
interfaces for database and web programming that are the basis
of the work presented in this paper.

Our framework and tool, called ``Spicey'', supports the following features:
\begin{description}
\item[ER-based:]
The framework is based on a specification of the data model
as an entity-relationship (ER) model. Thus, the complete source code
of an initial system is generated from an ER model.
\item[Web-based:]
The generated system is web-based, i.e., all data can be manipulated (i.e.,
created, shown, modified, deleted) via standard web browsers.
The initial system provides operations to insert new entities,
show entities, modify or delete existing entities
as specified in the ER model.
Relations between entities are manipulated together with
the corresponding entities. For instance, if there is a
one-to-many relation between $E$ and $E'$,
an instance of $E'$ can be created only if a corresponding instance
of $E$ is selected.
\item[Typed:]
The source code is statically typed so that many programming errors are
detected at compile time (in contrast to applications implemented
in Perl, PHP, Ruby, etc). Moreover, the data types
specified in the ER model are also respected, i.e., it is not
possible to submit web forms containing ill-typed data
so that the integrity of the stored data might be destroyed.
\item[Sessions:]
Since HTTP is a stateless protocol,
our framework provides a session concept so that any kind
of data (e.g., the contents of a virtual shopping basket)
can be stored in a user session.
Sessions are also used to store login information or
navigate the user through a sequence of interactions.
\item[Authentication:]
The generated application contains an initial structure
for authentication, i.e., login/logout operations.
Since the concrete authentication methods usually depend
on the application (e.g., kind of login names, passwords),
this initial structure must be extended by the programmer.
\item[Authorization:]
The generated application has methods for authorization,
i.e., each controller that is responsible for showing or modifying
data is authorized before execution.
A central authorization module is generated where the programmer
can easily specify authorization rules based on login or similar information.
\item[User processes:]
Individual operations provided by the framework can be
composed to user processes that can be selected to initiate longer
interaction sequences. For instance, if it is necessary
to create various entities in a database, the individual ``create''
operations can be connected to a complex user process.
A user process can be considered as a wizard-like dialog
spanning over multiple pages.
Such processes are specified as graphs using functional logic
programming techniques.
\item[Routing:]
As often found in complex web-based systems,
the routes (i.e., URLs to call some functionality of the system)
are decoupled from the physical structure of the source code.
This enables simple URLs and bookmarking of URLs that persist restructurings
of the implementation.
Therefore, our framework generates applications that contain
a specification of a mapping from URLs into controllers of the application.
\end{description}
In the remainder of the paper, we present the main features of our framework
and show how declarative programming is useful to get a
compact and maintainable implementation of web-based applications.
In the next section, we briefly survey Curry and its features
for web programming as required in this paper.
Section~\ref{sec:er} reviews the use of entity-relationship models
for database programming in Curry.
The generation of the basic structure of a web application
from an ER model is discussed in Section~\ref{sec:scaff}.
The remaining sections discuss the implementation of sessions,
authentication, authorization, and user processes
before we conclude in
Section~\ref{sec:conclusions} with a discussion of related work.

\section{Web Programming with Curry}
\label{sec:curry}

We briefly survey the basic concepts of Curry and their use
for high-level web programming as required to understand the main
part of this paper.
More details of Curry can be found in recent surveys on
functional logic programming \cite{AntoyHanus10CACM,Hanus07ICLP}
and in the definition of Curry \cite{Hanus12Curry}.

The design of the declarative multi-paradigm language Curry
is an attempt to integrate the most important features
of functional and logic languages in a seamless way
to provide a variety of programming concepts to the programmer.
Conceptually, Curry combines demand-driven evaluation,
parametric polymorphism, and
higher-order functions from functional programming
with logic programming features like
computing with partial information (logic variables),
unification, and non-deterministic search for solutions.
As shown in previous works on database programming
\cite{BrasselHanusMueller08PADL,Fischer05}
and web programming \cite{Hanus01PADL,Hanus06PPDP,Hanus07PPDP},
this combination enables better abstractions in application programs.
Curry has a Haskell-like syntax\footnote{%
Variables and function names usually
start with lowercase letters and the names of type and data constructors
start with an uppercase letter. The application of $f$
to $e$ is denoted by juxtaposition (``$f~e$'').}
\cite{PeytonJones03Haskell}
extended by the possible inclusion of free (logic)
variables in conditions and right-hand sides of defining rules.
The operational semantics of Curry, described in detail in
\cite{Hanus97POPL,Hanus12Curry}, is based on an optimal evaluation strategy
\cite{AntoyEchahedHanus00JACM} which is a conservative extension
of lazy functional programming and (concurrent) logic programming.
Curry also offers standard features of
functional languages, like modules or monadic I/O
which is identical to Haskell's I/O concept \cite{Wadler97}.
Thus, \ccode{IO $\alpha$} denotes the type of an I/O action that returns values
of type $\alpha$.

As a simple example for a Curry program,
consider the following data declarations.
The first declaration introduces a data type \ccode{Maybe a}
of possible values (of an arbitrary type \code{a}),
where \ccode{Nothing} is a constructor denoting the absent of a value
and the constructor \ccode{Just} decorates a given value.
The second declaration introduces a type \code{HtmlExp}
to represent HTML structures:
\startprog
data Maybe a = Nothing | Just a
data HtmlExp = HtmlText   String                              
             | HtmlStruct String [(String,String)] [HtmlExp]
\stopprog
Thus, an HTML expression is either a plain string (\code{HtmlText})
or a structure (\code{HtmlStruct}) consisting of a tag
(e.g., \ccode{b}, \ccode{em}, \ccode{h1}, \ccode{h2},\ldots),
a list of attributes (name/value pairs),
and a list of HTML expressions contained in this structure.
Since it is tedious to write HTML documents in this form,
we define various functions as useful abbreviations, like
\startprog
htxt   s     = HtmlText   (htmlQuote s)
par    hexps = HtmlStruct "p" [] hexps
italic hexps = HtmlStruct "i" [] hexps
\ldots
\stopprog
Then we can write HTML expressions like
\startprog
par [htxt "This is an ", italic [htxt "example"]]
\stopprog
As an example for an operation on HTML expressions,
we define a function \code{textOf}
that extracts the textual contents of an HTML structure
based on the predefined list processing operations \code{concat}
(to concatenate a list of lists) and \code{map}
(to apply an operation to every element of a list):
\startprog
textOf :: HtmlExp -> String
textOf (HtmlText s) = s
textOf (HtmlStruct t as hs) = concat (map textOf hs)
\stopprog
A \emph{dynamic web page} is an HTML document (with header information)
that is computed by a program at the time when the page is requested
by a client (e.g., a web browser).
Dynamic web pages usually process user inputs, placed in various
input elements (e.g., text fields, text areas, check boxes)
of an HTML form, in order to generate a user-specific result.
For this purpose, the HTML library of Curry \cite{Hanus01PADL}
provides an abstract programming model that
can be characterized as \emph{programming with call-back functions}.
A web page with user input and buttons for submitting the input to
a web server is modeled by attaching an \emph{event handler}
to each submit button that is responsible for computing the answer document.
For instance, the HTML library defines an operation to represent
submit buttons in an HTML page:
\startprog
button :: String -> HtmlHandler -> HtmlExp
\stopprog
In order to access the user input, the event handler
(of type \code{HtmlHandler}) has an environment containing
the actual user input as a parameter and computes a new web page.
We omit further details here, which can be found in \cite{Hanus01PADL},
since our framework is mainly based on a more abstract layer
to construct \emph{web user interfaces} (\emph{WUI}s) \cite{Hanus06PPDP}.
Such WUIs are constructed in a type-oriented manner,
i.e., for each type in the application program one can construct
a WUI that is an implementation of a web-based interface
to manipulate values of this type. Thus, the (tedious)
code for checking the validity of values in the input fields and
providing appropriate error messages is automatically derived
from the WUI specification. For instance, the corresponding WUI
library \cite{Hanus06PPDP} contains predefined WUIs to manipulate strings
(\code{wString}) or to select a value (\code{wSelect}) from a given
list of values (where the first argument shows a value as a string):
\startprog
wString :: WuiSpec String
wSelect :: (a -> String) -> [a] -> WuiSpec a
\stopprog
Here, \ccode{WuiSpec\,\,a} denotes the type of a WUI
to modify values of type \code{a}.
To construct WUIs for complex data types, there are
\emph{WUI combinators} that are mappings from simpler WUIs to WUIs
for structured types. 
For instance, there is a family of WUI combinators for tuple types:
\startprog
wPair   :: WuiSpec\,\,a -> WuiSpec\,\,b -> WuiSpec\,\,(a,b)
wTriple :: WuiSpec\,\,a -> WuiSpec\,\,b -> WuiSpec\,\,c -> WuiSpec\,\,(a,b,c)
w4Tuple :: WuiSpec\,\,a -> WuiSpec\,\,b -> WuiSpec\,\,c -> WuiSpec\,\,d
        -> WuiSpec\,\,(a,b,c,d)
\ldots
\stopprog
Hence,
\startprog
wPair wString (wSelect show [1..100])
\stopprog
defines a WUI to manipulate a pair of a string and a number between 1 and 100.
An important feature of WUIs is their easy adaptation to specific requirements.
For instance, there is an operator \code{withCondition}
that combines a WUI and a predicate on values so that the resulting
WUI accepts only values satisfying this predicate. Thus,
\startprog
wRequiredString = wString `withCondition` (not . null)
\stopprog
defines a WUI that accepts only non-empty strings.
Similarly, there are combinators to change the default rendering
of WUIs (\code{withRendering}) or to change the default error messages.
These features allow a compact and declarative description
of complex user interfaces.

We want to remark that the functional as well as logic features
of Curry are exploited to implement this high-level abstraction:
event handlers and environments are functions attached
to data structures representing HTML documents,
and input elements in a document have logic variables as references.
Moreover, static type checking is exploited to ensure type-safe web forms.

\section{Entity-Relationship Models and Database Programming}
\label{sec:er}

The entity-relationship model \cite{Chen76} is an established framework
to specify the structure and specific constraints of data stored
in a database. It is often used with
a graphical notation, called entity-relationship diagrams (ERDs),
to visualize the conceptual model.
The ER framework proposes to model the part of the world that is
interesting for the application by entities that have attributes
and relationships between the entities.
The relationships have cardinality constraints
that must be satisfied in each valid state of the database, e.g.,
after each transaction.

\citeN{BrasselHanusMueller08PADL}
developed a technique to generate high-level and safe database
operations (i.e., the cardinality constraints of the ER model
hold after database updates) from a given ERD.
In order to be largely independent of a specific ER modeling tool,
\citeN{BrasselHanusMueller08PADL} defined a representation of ERDs
in Curry so that graphical modeling tools can be connected
by implementing a translator from the tool format into
the Curry representation.
Since this representation is also the starting point of our framework,
we briefly describe it in the following.

If the structure of possible ERDs is fixed (unfortunately,
there is no standard definition of ERDs),
the representation of ERDs as data types in Curry is straightforward.
Here we assume that an ERD consists of a name
(that is later used as the module name
containing the generated database operations)
and lists of entities and relationships:
\startprog
data ERD = ERD String [Entity] [Relationship]
\stopprog
Instead of showing the detailed definition of all ER data types,
which can be found in \cite{BrasselHanusMueller08PADL},
we show the ER specification of an example which we use throughout
this paper: a web log. The structure of our ``blog'' is visualized
as an ERD in Fig.~\ref{fig:blogerd}.
A blog consists of \code{Entry} articles
having title, text, author, and date as attributes, and
\code{Comment}s to each entry. Furthermore, there are a number
of \code{Tag}s to classify \code{Entry} articles.
One can translate this ERD into the following data term
which specifies the details of the blog structure:
\begin{figure}[t]
\begin{center}
  \includegraphics[scale=0.7]{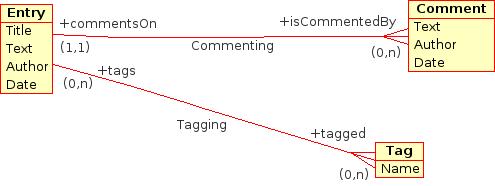}
\end{center}\vspace{-3ex}
\caption{An ER diagram of a web log}
\label{fig:blogerd}
\end{figure}
\startprog
ERD "Blog"
    [Entity "Entry"
       [Attribute "Title"  (StringDom Nothing) Unique False,
        Attribute "Text"   (StringDom Nothing) NoKey  False,
        Attribute "Author" (StringDom Nothing) NoKey  False,
        Attribute "Date"   (DateDom   Nothing) NoKey  False],
     Entity "Comment"
       [Attribute "Text"   (StringDom Nothing) NoKey  False,
        Attribute "Author" (StringDom Nothing) NoKey  False,
        Attribute "Date"   (DateDom   Nothing) NoKey  False],
     Entity "Tag"
       [Attribute "Name"   (StringDom Nothing) Unique False] ]
    [Relationship "Commenting"
       [REnd "Entry"   "commentsOn"    (Exactly 1),
        REnd "Comment" "isCommentedBy" (Between 0 Infinite)],
     Relationship "Tagging"
       [REnd "Entry" "tags" (Between 0 Infinite),
        REnd "Tag" "tagged" (Between 0 Infinite)]            ]
\stopprog
Each attribute specification consists of the attribute name,
the domain type of the attribute values together with a possible
default value, and specifications
of the key and null value property.
For instance, the \code{Title} attribute of the entity \code{Entry}
is a string without a default value, specified by
\ccode{(StringDom Nothing)},
 that is unique in each valid state of the database,
and null values are not allowed for this attribute.
Furthermore, \code{Commenting} is a one-to-many relationship between
\code{Entry} and \code{Comment} entities
(\ccode{(Exactly 1)} denotes the interval $[1..1]$
and \ccode{(Between 0 Infinite)} denotes the interval $[1..\infty]$).
Hence, each \code{Entry}
article has an arbitrary number of comments and each \code{Comment}
belongs to exactly one \code{Entry}. Finally,
\code{Tagging} is a many-to-many relationship between \code{Entry} and
\code{Tag} entities.

As mentioned above, \citeN{BrasselHanusMueller08PADL}
proposed a method to generate database operations from
an ERD specification that ensures the integrity of the database
(w.r.t.\ the constraints present in the ERD)
after performing update operations. For instance, there is
an operation of type
\startprog
newEntry :: String -> String -> String -> CalendarTime
         -> Transaction Entry
\stopprog
that takes values of the \code{Entry} attributes and inserts
a new \code{Entry} entity into the database.
The return type is a transaction (see \cite{BrasselHanusMueller08PADL}),
i.e., the insertion might fail (without changing the database state
but returning some informative error message)
if the value of the title attribute is not unique.
Similarly, there is a generated operation of type
\startprog
newCommentWithEntryCommentingKey
  :: String -> String -> CalendarTime -> EntryKey
  -> Transaction Comment
\stopprog
that takes values of the attributes of a new \code{Comment} entry
\emph{and} a key of an existing \code{Entry} entity since each
comment is related to a unique \code{Entry} entity,
as specified by the \code{Commenting} relation.

The main idea of our tool Spicey, described in the following
sections, is the generation of
a maintainable and adaptable web application that implements a user-friendly
interface to these database operations.

It should be noted that
the underlying database library is based on logic programming
techniques where the logic features of the language Curry
are exploited to embed a declarative query language into Curry,
as shown in \cite{BrasselHanusMueller08PADL,Fischer05}.
For this purpose, each database entity is represented as a predicate
between its database key and the corresponding entity instance
and each relationship of the ERD is represented as a predicate
between the corresponding database keys.
For instance, \ccode{comment ckey cmt} is satisfied
if \code{cmt} is a \code{Comment} instance with key \code{ckey},
and \ccode{commenting ekey ckey} is satisfied if the
\code{Entry} instance with key \code{ekey} is related to the
\code{Comment} instance with key \code{ckey} w.r.t.\ the relationship
\code{Commenting}.
Thus, we can join these predicates to obtain a query
that returns all comments belonging to a given entry key:
\label{sec:queryCommentsOfEntry}
\startprog
queryCommentsOfEntry :: EntryKey -> Query [Comment]
queryCommentsOfEntry ek =
  queryAll (\bs{}c -> let\,\,ck\,\,free\,\,in comment ck c <> commenting ek ck)
\stopprog
Here, \ccode{<>} denotes the join of two predicates,
the free variable \code{ck} denotes an arbitrary \code{Comment} key,
and \code{queryAll} is a query that returns all
solutions to a predicate abstraction.
More details can be found in \cite{BrasselHanusMueller08PADL}.

The advantages of the integration of database querying
into the programming language instead of using a decoupled
abstraction like SQL are type-safety, the possibility to use all
language features the programmer is used to, and the prevention
of security risks that might be introduced by
a string-based SQL interface \cite{Huseby03}.
Thus, the use of a logic-oriented implementation language
is essential to obtain our design, described below,
although the application of the logic features
are hidden by the database abstractions sketched in this section.

\begin{figure}[t]
\begin{center}
  \includegraphics[scale=0.49]{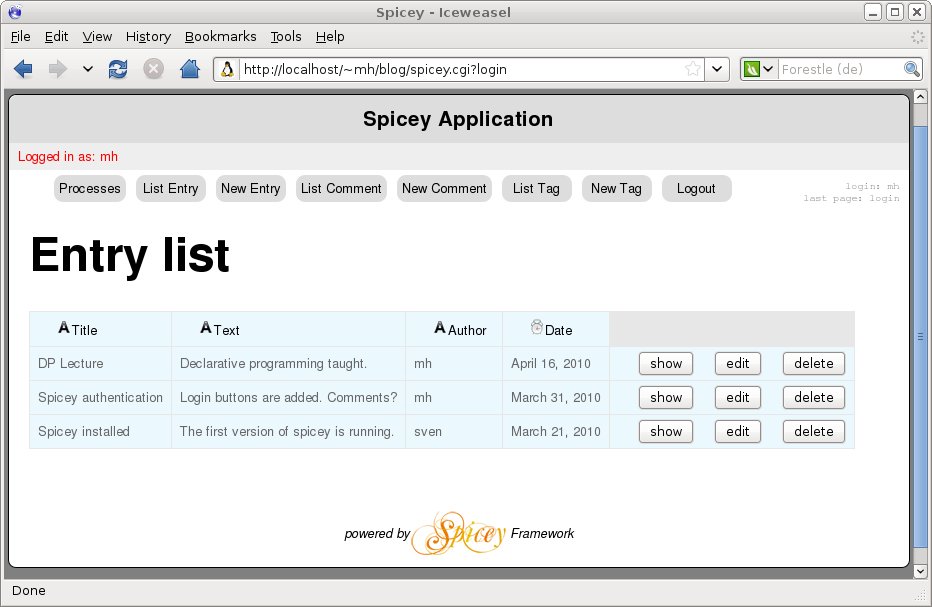}
\end{center}\vspace{-3ex}
\caption{The web interface of the blog application generated by Spicey}
\label{fig:spiceymain}
\end{figure}

\section{Scaffolding}
\label{sec:scaff}

In this section, we present the basic scaffolding of Spicey,
i.e., the generation of an initial executable system
that provides access to the data via standard web browsers.
In order to make the generated system maintainable,
it is important that the program code has a comprehensible structure.
Therefore, Spicey uses a
well-established code structure (also called pattern) for interactive systems:
the model-view-controller (MVC) structure \cite{KrasnerPope88}.
This is based on the idea to distribute the entire
functionality of an interactive system into three parts:
the \emph{model} which represents
the application data and contains all operations to manipulate these data,
the \emph{view} that is responsible to represent the model to the user,
and the \emph{controller} that reacts to user requests
and initiates changes in the model (and, thus, in the view).
Due to the diversity of data represented by the various
entities, Spicey generates various views and controllers
from a given ER model.
Before presenting more details of this scaffolding process,
we discuss some design decisions.

\subsection{Structure of Generated Applications}

As an example, consider the ER description of the blog
presented in the previous section.
{}From this description, Spicey automatically generates
the Curry source code of an application that implements
the interface shown in Fig.~\ref{fig:spiceymain}.
As illustrated, the interface has buttons to create
new entities and list existing ones, as well as buttons
to show, edit, or delete any existing entity.

However, generating a standard interface is not sufficient
for real applications since there are many requirements
that are not present in the ER description.
For instance, one might want to choose a different table layout
or show only the first 30 characters of the \code{Text}
attribute in the list of entries.
One could extend the ER descriptions to add specifications
of these requirements, but there are so many of these
requirements in real applications so that this leads to
a complex specification structure that is difficult to manage.
As an alternative, we propose to use the high abstraction level
of declarative programming for this purpose.
Instead of adding all possible customer requirement
to the specification language of the data model,
we generate high-level declarative code from the ER descriptions.
Thanks to the works on high-level database programming
and web user interface construction sketched above,
the generated source code is compact and comprehensible
so that it can be easily adapted to individual customer requirements,
as demonstrated below.

As mentioned above, the scaffolding of Spicey is based
on the model-view-controller structure
for the generated source code. The MVC structure is reflected
in the module structure of the code.
Thus, if we execute Spicey
to generate a web application from an ER description,
the following directories and modules are created:
\begin{description}
\item[\code{models/}] This directory contains the implementation
of the data model, i.e., it contains the Curry module
implementing the access to the database which is
generated from the ER description as sketched in Section~\ref{sec:er}
and described in detail in \cite{BrasselHanusMueller08PADL}.
In particular, this module contains, for each entity of the ER model,
a definition of an (abstract) data type representing such entities.
In our blog example, these are the data types \code{Entry},
\code{Comment}, and \code{Tag}.
If one wants to add more complex integrity constraints
on update operations for these entities, one could extend the Curry code
in this module.
\item[\code{controllers/}]
This directory contains the implementation of the various
controllers that are responsible to react on user interactions.
Some if these controllers can be directly called, e.g.,
from the main menu shown at the top of Fig.~\ref{fig:spiceymain},
whereas other controllers (e.g., for editing or deleting entities)
are called as continuations from particular views.
The general type of a controller in Spicey is simply
\startprog
type Controller = IO [HtmlExp]
\stopprog
Thus, a controller is an I/O action that returns
an HTML document, the result shown to the user,
which is embedded into the standard page layout by the scheduler.
For each entity of the ER model, Spicey generates
a corresponding controller module containing the controllers
to list, create, edit, and delete such entities.
For instance, the controller to edit a given \code{Comment} entity
is defined with the type
\startprog
editCommentController :: Comment -> Controller
\stopprog
\item[\code{views/}]
This directory contains the implementation of the views
of the different entities, i.e., a view module is generated
for entity of the ER model. These views are called from
the corresponding controllers.
For instance, there are views to show, insert, or edit an entity,
as well as a view to list all entities.
\item[\code{config/}]
This directory contains modules to configure the overall access
to the functionality provided by the system.
For instance, it contains information about the routes,
i.e., the URLs supported by the system and their mapping
to individual controllers, and the definition of available user processes
(see Section~\ref{sec:procs}).
\end{description}
Furthermore, there are directories containing
global modules for session management, authentication etc (\code{system/}),
scripts to compile and install the system (\code{scripts/}),
and collections of images and style files used by the system (\code{public/}).
In the following, we explain some parts of the generated source
code in more detail (where we omit some minor aspects
compared to the concrete code in order to simplify the discussion).

\subsection{Views}

\begin{figure}[t]
\begin{center}
  \includegraphics[scale=0.49]{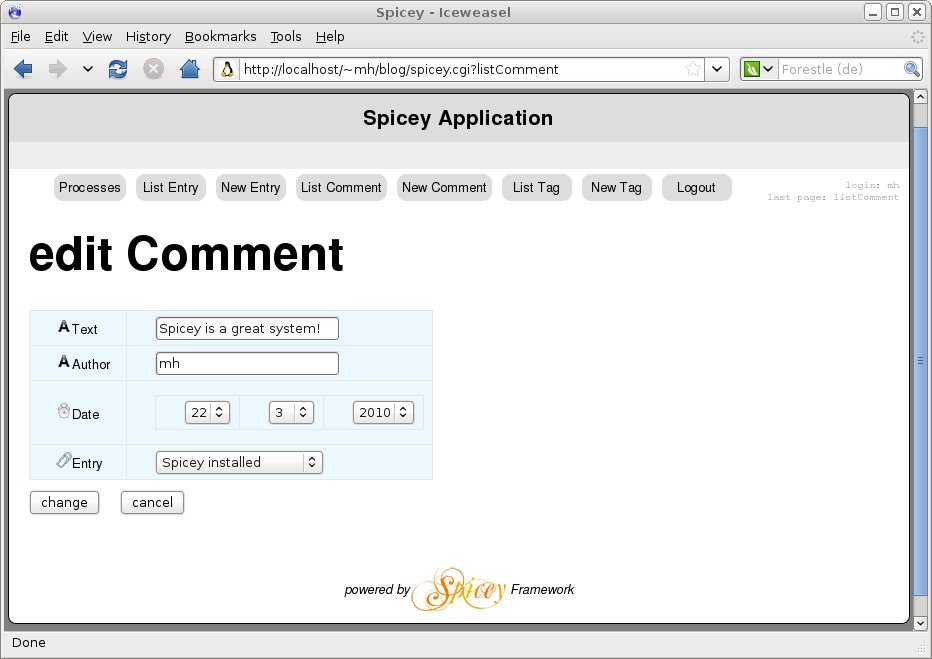}
\end{center}\vspace{-3ex}
\caption{An edit form for blog comments generated by Spicey}
\label{fig:spiceyedit}
\end{figure}

To obtain a compact and maintainable source code,
the \emph{views} that create or update entities exploit WUIs
(see Section~\ref{sec:curry}) to implement
type-safe web forms in a high-level declarative manner.
Thus, Spicey generates for each entity a WUI specification
of a web form to manipulate the attributes of this entity
(e.g., see Fig.~\ref{fig:spiceyedit}).
However, the internal primary database keys of an entity
should not be changed and, thus, they are not part of the
WUI specification. Moreover, if an entity is related to
other entities, this relation should be modifiable in the web form.
For instance, each comment in our blog example is related
to a unique \code{Entry} entity.
Hence, a single \code{Entry} entity must be selected in
the form to insert or change a comment
(see the lower selection box in Fig.~\ref{fig:spiceyedit}).
As a consequence, we have to pass related entities
to the web form in order to enable their selection.
In the generated code, we do not pass all associated
entities (e.g., it is not reasonable to select the associated
comments when editing an \code{Entry} entity)
but only the uniquely related entities from one-to-many relationships
and ``one side'' of many-to-many relationships.

To be more precise, assume that $E$ is
an entity with attributes $A_1,\ldots,A_n$,
$(E_1,E),\ldots,(E_k,E)$ are all one-to-many relationships (to $E$)
and $(E,E'_1),\ldots,(E,E'_l)$ are all many-to-many relationships
(with $E$ as the first component).
Then the form generated by Spicey to edit
an $E$ entity (as shown in Fig.~\ref{fig:spiceyedit} for a \code{Comment}
entity) contains the following components:
\begin{enumerate}
\item Input fields for editing the attributes $A_1,\ldots,A_n$
\item Selection fields to select the uniquely related entities $E_1,\ldots,E_k$
\item Multiple selection fields to select the related
entities $E'_1,\ldots,E'_l$
\end{enumerate}
Thus, one could select in our blog example
an \code{Entry} entity in a form to edit a \code{Comment} (due to
the one-to-many relationship \code{Commenting}) and
a set of \code{Tag} entities in a form to edit an \code{Entry}
(due to the many-to-many relationship \code{Tagging}).

Due to these considerations, Spicey generates from the \code{Blog} ERD
the following WUI specification for \code{Comment} entities:
\startprog
wComment :: [Entry] -> WuiSpec (String,String,CalendarTime,Entry)
wComment entries =
  (w4Tuple wRequiredString wRequiredString wDateType
           (wSelect entryToShortView entries))
    `withRendering` (renderLabels commentLabelList)
\stopprog
Thus, \code{wComment} takes a list of available entries
and returns a web form to manipulate the three attributes of
a \code{Comment} entity together with the uniquely associated
\code{Entry} entity.
The available entries are shown in a selection box (\code{wSelect})
where each entry is shown as a short string by the transformation
function \code{entryToShortView}.
As a default, the first unique attribute is used for this purpose
(if present), i.e., in case of an \code{Entry} entity,
the title of the corresponding entry is shown.

We want to remark that this and other defaults
used in the standard web form created by this
WUI specification (see Fig.~\ref{fig:spiceyedit})
can be easily adapted by changing this declaration.
For instance, one can use another interface for manipulating dates
by replacing \code{wDateType} with another WUI for dates,
or if the name of the author is not required
(i.e., if comments are accepted with an empty \code{Author} string),
one can replace the second \code{wRequiredString} by \code{wString}.
Moreover, the complete default rendering can be changed
by using another rendering function than \code{renderLabels}
(see \cite{Hanus06PPDP} for more details about the rendering).

The WUI operation \code{wComment} is used to implement the views
to insert or update a \code{Comment} entity.
For instance, for editing comments, Spicey generates an operation
\startprog
editCommentView
  :: Comment -> Entry -> [Entry] -> (Comment -> Controller)
  -> [HtmlExp]
\stopprog
that takes the current comment, the \code{Entry} entity related to this
comment, a list of available \code{Entry} entities, and
a controller to update the modified comment in the database as arguments.
Note that the \code{Comment} data type contains the foreign key
of the associated \code{Entry} entity so that it need not be explicitly passed
to the update operation, see also \cite{BrasselHanusMueller08PADL}.

The main view to browse and manipulate entities is the list view
as shown in Fig~\ref{fig:spiceymain}.
Since the list view contains buttons (show/edit/delete)
associated to individual entities,
the controllers implementing the functionality of these buttons
are passed as arguments to the view.
For instance, the implementation of the generated list view for
\code{Comment} entities is quite simple by the use of the
\code{HTML} library:
\startprog
listCommentView :: [Comment]
                -> (Comment -> Controller)
                -> (Comment -> Controller)
                -> (Comment -> Controller) -> [HtmlExp]
listCommentView comments showctrl editctrl deletectrl =
  [h1 [htxt "Comment list"],
   table ([take 3 commentLabelList] ++
          map listComment (sort leqComment comments))]
 where listComment cmt = commentToListView cmt ++
          [[button "show"   (nextController (showctrl   cmt)),
            button "edit"   (nextController (editctrl   cmt)),
            button "delete" (nextController (deletectrl cmt))]]
\stopprog
The list view has the list of comments and the necessary controllers
(\code{showctrl}, \code{editctrl}, \code{deletectrl})
as arguments and creates a table of comments and buttons having
the controllers as continuations. \code{nextController} is a global
operation which wraps the output of a controller with the standard
layout of the application.
The comments are sorted w.r.t.\ the ordering \code{leqComment},
an operation generated by Spicey. Thus, the generated default ordering
(a lexicographic ordering on the attributes of the entity)
can be easily changed.

To influence the information shown in the list view,
one has to adapt the definition of the generated
operation \code{commentToListView} which maps a \code{Comment}
entity into a row of the table.
The initial definition is simply the text of all attributes.
Spicey generates the definition of the various entity representations
used in the application, like short views, list views, or views containing
all details,
in single module (named \code{BlogEntitiesToHtml}).
Thus, one needs to adapt only this module to change the
default layout of the entities. This module also contains
the definition of the labels corresponding to the attribute names,
like the constant \code{commentLabelList} used in the list view and the
edit form.

\subsection{Controllers}

Following the MVC paradigm, \emph{controllers} are responsible
to react to user requests and call the corresponding views
supplied with data contained in the model.
For instance, the list controller for comments
retrieves all comments from the model (i.e., the database)
and calls the operation \code{listCommentView}
with these comments and the controllers to process individual comments:
\label{ex:listcommentcontroller}
\startprog
listCommentController :: Controller
listCommentController = do
  comments <- runQ (queryAll (\bs{}c->let\,\,key\,\,free in comment\,\,key\,\,c))
  return (listCommentView comments
                          showCommentController
                          editCommentController
                          deleteCommentController)
\stopprog
In order to implement the listing of a restricted set of comments
(e.g., all comments of a particular author), one can use in the
controller's code the operation
\startprog
getControllerParams :: IO [String]
\stopprog
that returns the parameters passed with the controller's URL.
For instance, one can easily define a controller for comments
that lists only the comments belonging to a given entity
(instead of listing all comments) by using the query
\code{queryCommentsOfEntry} shown in Section~\ref{sec:queryCommentsOfEntry}.

The other controllers are similarly defined. However, note that
controllers to create or modify entities require a second
controller, passed to the view (e.g., see \code{editCommentView} above),
that is responsible to perform the actual modification of the model.
All controllers for an entity generated by Spicey are put
into a module, e.g., the module \code{CommentController}
contains the various controllers associated to \code{Comment} entities.

\subsection{Routing}

As shown in Fig.~\ref{fig:spiceymain},
some controllers (like \code{new} or \code{list})
can be directly called by specific URLs in the application.
In order to decouple the structure of URLs from the structure
of the implementation (which is reasonable to hide its details),
Spicey generates an initial module containing the names of the
available controllers and their URLs.
An indirection in this generation is necessary due to
potential cyclic module dependencies which are not allowed in Curry.
Controller modules depend on view modules since controllers call
view operations. If one wants to put in some view also
URL references to controllers, we obtain a cyclic dependency.
Therefore, Spicey generates a data type that enumerates all
``top-level'' controllers, i.e., controllers that can be activated
by URLs:
\startprog
data ControllerReference = ListEntryController
                         | NewEntryController
                         | ListCommentController
                         | \ldots
\stopprog
The mapping of these controller references to the actual
controller operations is defined in a top-level module that
is used only by the main module of the application (this avoids
the cyclic dependency).

The routing, i.e., the association of URLs and controllers, is defined
by an operation \code{getRoutes} that is initially defined as follows
(we omit the processes and login controllers since they are later discussed):
\startprog
getRoutes =
  return [("new Entry", Exact "newEntry", NewEntryController),
          ("list Entry",Exact "listEntry",ListEntryController),
          \ldots
          ("default", Always, ListEntryController)]
\stopprog
The first argument of each route element is the name as shown in
the top menu of the application (see Fig.~\ref{fig:spiceymain}),
the second argument specifies the matching of a route name as used in the URL
(where \code{Exact} defines an exact matching, \code{Always} defines
an always successful matching, and there is also an option
to define arbitrary matching functions),
and the third argument is the controller reference associated to the
matched URL.
In the default configuration, the top-level menu of the application is
dynamically generated from the \code{Exact} matchings defined
in \code{getRoutes}.

Altogether, a Spicey application performs a request for a web page as follows.
First, the path component of the URL is extracted.
Then, a dispatcher matches this path against the list of alternatives
defined by \code{getRoutes} and the controller reference of the first matching
alternative (or an error message controller if there
is no matching alternative) is returned.
Finally, the top-level module executes the code associated to
this controller reference and decorates the computed HTML contents
with the standard layout of the application.

Note that \code{getRoutes} is an I/O operation rather than a constant.
This allows a dynamic routing depending on some state of the
system. For instance, the available routes can be restricted
for users that are not logged in, or different routes
can be supported depending on the login status.
The implementation of these features requires
the management of sessions which is discussed in the next section.

\section{Sessions}
\label{sec:sessions}

In a web-based application, one needs a concept of a \emph{session}
in order to pass information between different web pages.
For instance, the login name of a user or the contents
of a virtual shopping basket should be stored
across several web pages.
Therefore, Spicey supports a general concept to store arbitrary
information in a user session.

Typically, sessions are implemented in web-based systems
via cookies stored in the client's browser.
For security and performance reasons,
these cookies should not contain the information
stored in the session but only a unique session identifier
that is passed to the web server in any interaction.
Therefore, a Spicey application implements sessions
by managing a \emph{session identifier} of the abstract type
\code{SessionID} in each web page.
If a session identifier does not exist (i.e., the browser did not send
a corresponding cookie), a fresh session identifier
is created and stored in a cookie sent with any subsequent web page.
This access to the current session identifier is implemented
in an operation
\startprog
getSessionId :: IO SessionId
\stopprog
However, the application programmer must not use this internal
operation to store session information.
Instead, Spicey provides the following operations to manipulate
session information (where the type variable \code{a} denotes
the type of the session information):
\startprog
getSessionData    :: Global (SessionStore a) -> IO (Maybe a)
putSessionData    :: a -> Global (SessionStore a) -> IO ()
removeSessionData :: Global (SessionStore a) -> IO ()
\stopprog
\code{getSessionData} retrieves information of the current session
(and returns \code{Nothing} if there is no information stored),
\code{putSessionData} stores information in the current session,
and \code{removeSessionData} removes such information.
\ccode{SessionStore a} is an abstract type to represent session information
containing data of type \code{a}.
This interface is based on the concept of ``globals''
(available through the Curry library \code{Global}\footnote{%
\url{http://www.informatik.uni-kiel.de/~pakcs/lib/CDOC/Global.html}})
that implements objects having a globally declared name in some module
of the program.
The values associated to the name can be modified by I/O actions.
It is also possible to declare global entities as persistent
so that their values are kept across different program executions,
but this is not required here since there is one process on the server
side serving all requests of a user session.

For instance, consider the implementation of ``page messages''
that are shown in the next page (e.g., error messages, status information),
like the \ccode{Logged in as} message shown in Fig.~\ref{fig:spiceymain}.
In order to enable the setting of such messages in any part
of a Spicey application, we define the page message as session data
by the following definition of a global entity:
\startprog
pageMessage :: Global (SessionStore String)
pageMessage = global emptySessionStore Temporary
\stopprog
\ccode{global $v$ Temporary} denotes a global entity with initial value
$v$ that is not persistently stored.
The value \code{emptySessionStore} denotes a session store that does
not contain any information.

Using the session operations above, we can define an operation
to set the page message in any part of a Spicey application:
\startprog
setPageMessage :: String -> IO ()
setPageMessage msg = putSessionData msg pageMessage
\stopprog
The current page message is retrieved and then removed by the following
operation:
\startprog
getPageMessage :: IO String
getPageMessage = do
  msg <- getSessionData pageMessage
  removeSessionData pageMessage
  return (maybe "" id msg)
\stopprog
This operation can be used by the main operation that wraps
a view output with the standard layout containing the page message,
global menu etc.

As one can see, the management of sessions using cookies
and session identifiers is completely hidden for the
application programmer.
The implementation of the operations to manipulate session data
is quite easy using session identifiers and appropriate
data structures.
For instance, the type \code{SessionStore} is implemented as a list
\startprog
data SessionStore a = SStore [(SessionId, ClockTime, a)]
\stopprog
where each element consists of a session identifier,
a clock time value (used to clean up the store from old data),
and the associated session data.
Then, the implementation of the operation
\code{getSessionData} amounts to a lookup of the information
associated to the current session identifier in the global session store,
or \code{putSessionData} simply adds or updates this information.

Due to this general session concept, one can easily attach
any number of information entities to a session.
For instance, one can store the history of selected controllers
(to implement a history list or a ``back'' button)
or the login name in order to support authentication,
which is discussed next.

\section{Authentication and Authorization}
\label{sec:auth}

The basic support for user authentication is quite simple.
One can define some session data to store a login name:
\startprog
sessionLogin :: Global (SessionStore String)
sessionLogin = global emptySessionStore Temporary
\stopprog
and use the session data operations to set, retrieve, or delete
a login name.
These operations can be used in specific web pages to login
or logout. Since authentication is required in almost any
web-based system keeping some data, Spicey provides
an initial implementation (compare Fig.~\ref{fig:spiceymain})
that is intended for extension
during the adaption of the system.
Although the initial authentication system is incomplete
(since it is not specified where to store passwords, login names etc),
its implementation provides a reasonable structure
that can be extended by the application programmer.
Moreover, the generated Spicey application also contains
some useful operations to generate random passwords,
compute hash strings for passwords and login names (note
that, for security reasons, one should not hash passwords alone
\cite{Huseby03}), etc.

An equally important aspect of web-based systems is authorization,
i.e., the checking whether a user is allowed to call
a distinct functionality, like showing or updating particular
entities.
In our framework, this check can be performed before starting
a controller.
In order to avoid the distribution of these checks over
the entire implementation and keep the authorization rules
at a centralized place, Spicey decorates the generated code
of each controller with a call to some authorization code.
For this purpose, there is a data type
\startprog
data AccessResult = AccessGranted | AccessDenied String
\stopprog
and an operation
\startprog
checkAuthorization :: IO AccessResult -> Controller -> Controller
\stopprog
which takes an I/O operation for authorization checking
(returning an \code{AccessResult}) and a controller as arguments.
If the authorization returns \code{AccessGranted}, the controller
is executed, otherwise an error message is displayed.
In order to define concrete authorization rules for the
various controllers, Spicey generates a data type to classify
the controllers:
\startprog
data AccessType a = NewEntity | ListEntities | ShowEntity a 
                  | UpdateEntity a | DeleteEntity a
\stopprog
Now, the execution of each controller is protected by adding
an authorization check to the controller's code.
For instance, the generated code of the controller to list all
\code{Comment} entities (see Section~\ref{ex:listcommentcontroller})
is extended as follows:
\startprog
listCommentController =
  checkAuthorization (commentOperationAllowed ListEntities)
                     (do comments <- runQ \ldots
                         \ldots )
\stopprog
Thus, the actual authorization rules are collected in a single
module containing the definition of all operations used in
the calls to \code{checkAuthorization}. For instance, the default
definition of \code{commentOperationAllowed} is
\startprog
commentOperationAllowed :: AccessType Comment -> IO AccessResult
commentOperationAllowed \us{} = return AccessGranted
\stopprog
authorizing all \code{Comment} operations.
By refining this definition, one can specify restrictions
on the controllers depending on the various operations, specific
entities, or login information of the user.
For instance, a generic policy that disallows delete operations
can be expressed as follows:
\startprog
disallowDelete at = case at of
  DeleteEntity \us{} -> return (AccessDenied "Delete not allowed!")
  \us{}              -> return AccessGranted
\stopprog
Note that the logic programming features of Curry
can be quite useful here to specify authorization policies
in a rule-oriented manner.

\section{Processes}
\label{sec:procs}

Web-based applications generated by Spicey support individual interactions
to insert, show, and change any entity.
If the data model is complex and consists of many entity types,
it might be necessary to combine single interactions
to longer interaction sequences.
For instance, if one wants to insert new data where
different entities are involved, it is reasonable
to define an interaction sequence where the controllers
to insert the various new entities are sequentially activated.
Thus, one wants to offer \emph{user processes} (which can be also
considered as parts of complex business processes)
that are structured compositions of elementary interactions.

In order to support the implementation of processes,
a Spicey application has an infrastructure to define and
execute such processes.
{}From an abstract point of view, a process is a sequence of
calls to controllers. Therefore, processes can be weaved
into the default structure of controllers.
For this purpose, each controller which terminates
an individual interaction has a ``continuation'' controller
that is called in the next step. For instance, a controller
responsible for creating a new entity calls the list controller
of the same entity type, as in the controller which adds a new
\code{Tag} entity:
\startprog
createTagController name = runT (newTag name) >>=
  either (\bs\us{}     -> nextInProcessOr listTagController Nothing)
         (\bs{}error -> displayError \ldots)
\stopprog
Thus, the execution (\code{runT}) of the transaction
\code{(newTag name)}, that should insert a new \code{Tag} name
into the database,
calls, if successful, the \code{listTagController},
or displays an error message if the transaction fails (e.g.,
since the new name already exists).
However, the next controller is not directly called but
indirectly through the operation \code{nextInProcessOr}.
This operation checks whether the system executes a user process.
If no process is active, the given controller is called,
otherwise the controller specified in the next process state
is executed. In order to make the selection of the next
process state dependent on some information provided
by the previous controller (this is useful to implement loops
or branches in processes), the second argument
of \code{nextInProcessOr} might contain such information.
Thus, the application programmer can replace the default
value \code{Nothing} by some information available in the previous
controller.

The concrete structure of processes is defined in
a distinguished module \code{UserProcesses} as data of the following type:
\startprog
data Processes st = ProcSpec [(String,st)]
                             (st\,\,->\,\,ControllerReference)
                             (st\,\,->\,\,Maybe\,\,ControllerResult\,\,->\,\,st)
\stopprog
The type parameter \code{st} is the type of the states of a process,
which could be a number or some more informative enumeration type.
Hence, a process specification consists of a list
of start states together with a textual description (these start
states can be selected in the process menu),
a mapping of each state into a corresponding controller to be
executed in this state, and a state transition function
that maps a state into a new state depending on some optional result
provided by the previous controller (the type of these results
is \code{ControllerResult}, which is identical to \code{String}
in the default case).

We can use all features available in Curry to define processes.
For instance, one can compute the next state in a process
based on solving constraints w.r.t.\ the data in the model.
In general, the state transition function is partial,
i.e., if a process state has no successor, the process will be terminated.
If a state has more than one successor, the first one is selected
(multiple successor states can occur in situations like the insertion
of several entities in an arbitrary order).

As a concrete example, consider a simple process to insert a new tag
followed by the creation of a new \code{Entry} entity and terminated
with showing the list of all tags.
If we use numbers as state identifiers, we can specify
this process as follows:
\startprog
let controllerOf 0 = NewTagController
    controllerOf 1 = NewEntryController
    controllerOf 2 = ListTagController
\smallskip
    next 0 \us{} = 1
    next 1 \us{} = 2
 in ProcSpec [("Insert new tag and entry",0)] controllerOf next
\stopprog
Since the next process state is always fixed and does not depend
on some data from the previous controller in this simple example,
the second argument of the state transition function \code{next}
is not relevant and, hence, ignored in the definition of \code{next}.
If this specification is contained in the module \code{UserProcesses},
the process can be selected and stepwise executed in the web application.

\section{Related Work}
\label{sec:related}

Although Spicey is the first web programming framework
for a declarative language based on ER models and with support
for typical requirements in the area (e.g., safe transactions,
sessions, authentication, authorization, processes),
there are many related approaches.
In the following, we discuss the relation of Spicey
to some other approaches.

In contrast to other systems implemented in scripting
languages like Perl, PHP, or Ruby,
our implementation is statically typed so that many programming errors
that easily occur in such complex systems are detected at compile time.
For instance, all input fields in the views (web pages)
are statically typed similarly to the attributes and access
operations for the underlying database. Thus, programming errors
that confuses this data can be detected at compile time.
Compared to Ruby on Rails, a framework with similar objectives,
Spicey can be considered as an approach to show that
declarative programming allows the compact construction
of web-based systems with static type checking
(thus, supporting programming safety)
without the need for (unreliable) dynamic meta-programming techniques.
Spicey also uses a functional logic abstraction
to databases which allows the formulation of queries
as typed expressions of the language Curry.
In contrast to our  approach, Ruby on Rails uses the Active Record Query
Interface as an abstraction for SQL which is still mostly string-based and,
therefore, introduces security risks.
In order to obtain these advantages of Spicey, some design difficulties
had to be solved, like avoiding mutual module dependencies
by passing continuation controllers to views, routing, etc.

The Web Application Maker\footnote{\tt\url{http://www.declarativa.com/wam/}}
(WAM) is a framework with similar goals to those of Spicey.
The WAM generates a web interface from the meta-data of a relational
database, allowing the interface to be adapted to
specific user requirements. In contrast to WAM,
Spicey uses ER models, which usually contain more structural information,
to generate the database schema \emph{and} the corresponding web interface.

The iData toolkit \cite{PlasmeijerAchten06FLOPS}
is a framework, implemented with generic programming techniques
in the functional language Clean, to construct type-safe web interfaces
to data that can be persistently stored.
In contrast to our framework, the construction of an application
is done by the programmer who defines the various iData elements,
where we generate the necessary code from an ER description.
Hence, integrity constraints expressed in the ER description
are automatically checked in contrast to the iData toolkit.

Turbinado\footnote{\tt\url{http://www.turbinado.org/}}
is a web framework for Haskell. It is based on similar ideas
as Ruby on Rails but exploits static type checking for
more reliable programming, similarly to Spicey.
In contrast to our framework, Turbinado supports scaffolding only
to implement an object-relational mapping of the models,
and it is not based on an ER specification to ensure
integrity constraints in the application.

Seam \cite{YuanOrshalickHeute09}
is a complex framework for developing enterprise applications in Java. It
integrates many other projects to support a wide range of technologies. The
database abstraction is provided by an Enterprise Java Beans 3.0 implementation,
Hibernate by default, which enables the programmer to generate the
database schema directly from the model classes. In contrast to the ERD library
used by Spicey, there is no graphical way to create the models of the
application. Another disadvantage of Seam is the absence of a
single place to define consistency rules for data. There are three
places where consistency and validation rules may be defined.
The first two are the code of the models and the generated database schema.
Some, but not all, rules which are defined in the models through annotations are
put into the database schema, but often the programmer has to assure database
consistency by himself.
Seam supports the definition of the standard relationship types one-to-one,
one-to-many, many-to-one and many-to-many but provides no good way to enforce
ranges for the multiplicity of those relationships as Spicey does.
For example, a one-to-one relationship does not ensure
that there is always an entity on the other side of the relation
but that there may be an entity or null. As a consequence, a programmer
in Seam has to check for the presence of an entity by himself.
Hibernate provides an annotation for that,
but it is not fully integrated into Seam yet. 
The third place to define validation rules are the views,
for which Seam uses Java Server Faces.
Rules defined in the model are not automatically reflected in the views, simple
validation rules like required fields have to be defined again in the view,
which leads to inconsistency if those rules for a model are defined differently
in different views. Seam integrates the
jBPM\footnote{\tt\url{http://www.jboss.com/products/jbpm/}} project for modeling
business processes. jBPM defines the process in XML format where a
graphical editor exists.
Similarly to Spicey, the coupling of the process with the code is achieved by
connecting controller methods with the process. For
authorization another tool may be used in Seam, namely JBoss
Rules\footnote{\tt\url{http://www.jboss.com/products/rules/}}, which
provides a logical language for defining authorization rules.
This aspect is directly integrated into Spicey by the logic programming
features of Curry.

The web framework Seaside\footnote{\tt\url{http://www.seaside.st/}}
is based on the object-oriented language Smalltalk.
Seaside is one of the few frameworks that use the
\emph{Transform-View} pattern for views. This enables the compiler to check
the integrity of the views because they are defined as program code
instead of HTML templates. Spicey uses the same approach but
provides for stronger code checks due to the static type system of Curry.
Seaside supports process modeling by providing a stateful environment
over multiple requests and enable the programmer to span a controller
method over more than one page. In contrast to Spicey, processes
are not decoupled from the controller logic so that a high abstraction
level of processes as in Spicey is not obtained.

Django\footnote{\tt\url{http://www.djangoproject.com/}}
is a popular web framework
for the language Python which has features very similar to Ruby on Rails. The
implementation of routes for Spicey was inspired by the way Django handles
routes. While Django offers only regular expressions for matching URLs, Spicey
generalizes this concept and supports arbitrary computable functions
for determining the controllers associated to URLs.

\section{Conclusions}
\label{sec:conclusions}

We have presented the tool Spicey to generate web applications for
data models that are specified as entity-relationship models.
Spicey enables the generation of a fully functional
system from an ER description in a few seconds.
The usefulness of this initial system goes beyond the evaluation of
the feasibility of the data model.
Due to the use of a declarative target language,
the generated code is compact and comprehensible so that
it can be easily extended and adapted
to specific customer requirements.
This has been also achieved by the use of previous works
on declarative database and web programming that
supports a compact executable description of web interfaces.
Furthermore, the system generated by Spicey has an infrastructure for
many aspects related to web-based systems, like transactions that are safe
w.r.t.\ the ER constraints, sessions, authentication,
authorization, user-oriented processes, or routing.

To get an idea of the size of the generated source code
that might be inspected by the application programmer to adapt
the initial system, we counted the lines of code of the
application generated for the \code{Blog} data model
shown in Section~\ref{sec:er}.
The generated views contain 300 lines of code,
the generated controllers contain 200 lines of code,
and the configuration files (e.g., routing, default authorization)
contain 65 lines of code.
Of course, the complete executable has much more code,
like system libraries, specific Spicey libraries,
generated database code etc.
However, this code is usually irrelevant when
adapting the system to specific layout requirements.
As usual in current web-based systems, many layout details
are specified in a global style sheet file so that
the views generate only the basic structure of each web page.

Spicey is completely implemented in Curry.
The implementation is freely available.\footnote{%
\tt\url{http://www.informatik.uni-kiel.de/~pakcs/spicey/}}
Apart from some example applications, Spicey has been used
to provide web-based interfaces to existing databases
by the definition of appropriate ER descriptions
and to implement a system to manage module descriptions and study programs
for university curricula.
The latter system is in daily use at the university of Kiel
and the ER-based generation of the high-level declarative code
was quite useful to adapt the system to ongoing user requirements.

For future work, it would be interesting to develop a concept
for migration, i.e., to support changes in the ER model
that might entail changes in the generated and possibly adapted
application code.
Furthermore, it would be useful to implement a tool that
allows to mix Curry code with HTML code fragments, e.g., as shown with
the Haskell Server Pages \cite{MeijervanVelzen00},
in order to allow an easier integration of layouts developed
by HTML designers into the application programs.

\paragraph{Acknowledgements.}
The authors are grateful to the anonymous referees
for helpful comments and suggestions.



\end{document}